\documentstyle[prl,aps,epsf,psfig,multicol]{revtex}
\begin{document}
\draft

\title{Plaquette Ground State in the Two-dimensional $SU(4)$ Spin-Orbital Model}
\author{Mathias van den Bossche$^1$,
Fu-Chun Zhang$^2$ and Fr\'ed\'eric Mila$^1$}
\address{$^1$ Laboratoire de Physique quantique,
Universit\'e Paul Sabatier, 31062 Toulouse Cedex - France\\
$^2$ Department of Physics, University of Cincinnati, Cincinnati,  OH 
45221-0011 - USA.}
\maketitle
\begin{abstract}
In order to understand the properties of Mott insulators with strong ground
state orbital fluctuations, we study the zero temperature properties of the
$SU(4)$ spin-orbital model on a square lattice. Exact diagonalizations of finite
clusters suggest that the ground state is disordered with a singlet-multiplet 
gap and possibly low-lying $SU(4)$ singlets in the gap. An interpretation in 
terms of plaquette 
$SU(4)$ singlets is proposed. The implications for LiNiO$_2$ are discussed.
\end{abstract}
\begin{multicols}{2}
\narrowtext
Orbital degeneracy is a very common feature of Mott insulators. In most
cases it is lifted by a cooperative Jahn-Teller distortion at relatively high
temperature, and the low energy physics can be described by a pure spin model,
with an effective Hamiltonian, hence a magnetic order, that depends on 
the orbital ordering\cite{kugel}.
In the past few years, this picture has been challenged in a number of systems,
and the possibility to get a spin liquid is now well 
established\cite{pen,feiner,santoro}. But it seems that there are even more exotic
systems that do not undergo a cooperative Jahn-Teller distortion in spite of the
orbital degeneracy.
The best example is probably LiNiO$_2$, in which no orbital or magnetic order
has been detected down to very low temperature\cite{kitaoka}. The minimal
model to describe
this system is the $SU(4)$ spin-orbital model defined by the Hamiltonian
\begin{equation}
H = J \sum_{<i,j>} (2\vec s_i.\vec s_j+\frac{1}{2}) 
(2\vec \tau_i.\vec \tau_j+\frac{1}{2})$$
\label{hamilt1}
\end{equation}
on the triangular lattice. In this model, $\vec s_i$ are spin-1/2 operators that
describe the spin degrees of freedom of Ni$^{3+}$, while $\vec \tau_i$ are
pseudo-spin-1/2 operators that describe the orbital degeneracy associated to the
two $e_g$ orbitals. While additionnal terms arising from the anisotropy in
hopping integrals and the Hund's rule coupling will destroy
the symmetry between spin and pseudo-spin and favour parallel
alignment of the spins of a pair of neighbouring sites, the absence of
ordering may be traced back to the properties of the Hamiltonian (\ref{hamilt1}).
As suggested by Li {\it et al}\cite{li}, the ground state is a liquid of
resonant plaquette $SU(4)$ singlets. Note that this model is not equivalent to
the model of Ref.\cite{santoro}. Although both models possess $SU(4)$
symmetry, the low-energy physics is completely different. Hamiltonian (\ref{hamilt1}) 
in 1-dimension (1D) has been solved by Bethe Ansatz \cite{Sutherland,li2},
by numerical mehods \cite{ueda,pati,frischmuth,mila1} and field theory methods
\cite{azaria}, and  the ground
state is a spin-orbital liquid.  In 2D, the proposed plaquette
ground state of model (\ref{hamilt1}) was mainly based on variational wavefunction or
mean field theory \cite{li}, and more work  is clearly needed to put these
ideas on a firm ground.

In this Letter, we present a detailed analysis of the low-energy properties of
the $SU(4)$ model on the square lattice using symmetry
analysis and exact diagonalizations of finite clusters.
While the most relevant compound LiNiO$_2$ is of quasi-two-dimensional 
system made out of triangular planes, we shall start with the square  lattice 
for simplicity.  We
will discuss possible differences between the triangular and the square
lattices  at the end. The choice of exact diagonalization as a numerical
method was motivated by the fact that other methods that have been
successfully used in the 1D case cannot be applied here. In particular, the
Quantum Monte Carlo algorithm used by Frischmuth {\it et al.}\cite{frischmuth}
suffers from a  severe minus sign problem in 2D lattices.

Let us start with some symmetry considerations that will be very
useful throughout the paper. First of all, the $SU(4)$ symmetry implies that 
$s^z=\sum_i s_i^z$, $\tau^z=\sum_i \tau_i^z$ and $s\tau^z=\sum_i s_i^z 
\tau_i^z$ are good
quantum numbers, and all numerical results have been obtained by diagonalizing
the Hamiltonian in sectors defined by a given set $(s^z,\tau^z,s\tau^z)$. 
Besides, 
the first Casimir operator, the equivalent of the square of the total spin in 
$SU(2)$, is given for $N$ sites by $C_{1...N}=(1/32) (\vec A_{\rm tot})^2$, where 
the components of $\vec A_{\rm tot}$ are 
the fifteen generators of the $SU(4)$ algebra and are given by $2\sum_{i=1}^N 
s_i^\alpha$,
$2\sum_{i=1}^N \tau_i^\alpha$ and $4\sum_{i=1}^N s_i^\alpha \tau_i^\beta$, 
$\alpha,\beta=x,y,z$. Since the Casimir operator of any irreducible
representation (IR) can be easily calculated with the tools of group
theory\cite{cornwell}, this operator is useful to find 
out to which IR a given state belongs. The values of $C$
for various IR's, classified according to their dimensionality $d$, are
listed in Table I. The Hamiltonian can 
also be
written in terms of on site fifteen-component vectors as
\begin{equation}
H = \frac{J}{4} \sum_{<i,j>} (\vec A_i.\vec A_j + 1)
\end{equation}
This allows to rewrite the Hamiltonian of several small systems in terms 
of the Casimir operators of sub-systems using identities such as
 $\vec A_i.\vec A_j = (1/2) (32 C_{ij}-\vec A_i^2-\vec A_j^2)$. 
In this case, all eigenvalues
and degeneracies can be deduced from the possible IR's for each sub-system. 
As we shall see below, this allows a full diagonalization for systems with 2 and
4 sites, as well as for 8 sites with periodic boundary conditions where the
the dimension of the Hilbert space is already 65,536.

\vspace{5mm}

\begin{tabular}{|l|l|l|l|l|l|l|l|l|l|l|}
\hline
\ $d$ \ & 1 \ & ~~~4 \ & ~6 \ & ~10 \ & 15 \ & 20 \ & 20 \ & ~~20\ & 35 \ & 45  \\
\hline
 \  $C$ \ & 0 \ & $15/32$ \ & $5/8$ \ & $9/8$ \ & 1 \ & $3/2$ \ & $3/2$ \ & $63/32$\ & 3 \ & 2  \\ 
\hline
\end{tabular}

\vspace{1mm}

{\small TABLE I.  Dimension ($d$) and Casimir operator eigenvalue ($C$) of some 
irreducible representations of $SU(4)$. Note that there are three different IR's
with dimension 20, two of them with the same Casimir eigenvalue (3/2). }

\vspace{3mm}

Let us now present the results we have obtained for several systems. Since the
interpretation we will give at the end of the paper heavily relies on the 
properties of the 2 and 4 site clusters, we include them here for convenience.

{\it i) Two sites (pair):} In terms of spins and orbitals, the ground state is 6-fold
degenerate (spin singlet $\times$ orbital triplet and ${\it vice versa}$) with
energy $-J$. The other 10 states are degenerate with energy $+J$. In $SU(4)$
language, this means that the only accessible IR's have dimension 6 and 10. So
it is impossible to build an $SU(4)$ singlet with only two sites, as already
emphasized by Li {\it et al.}\cite{li}.

{\it ii) Four sites (plaquette):} The Hamiltonian can be rewritten as
\begin{equation}
H = 4J (C_{1234}-C_{13}-C_{24}+1/4)
\end{equation}
The ground state is an $SU(4)$ singlet with the pairs (13) and (24) in the IR of
dimension 6, and its energy is $-4J$. It minimizes the energy per bond and is
thus a very stable object. The first excited state is 50-fold
degenerate with energy $-2J$. This corresponds to twice the adjoint IR of
dimension 15 with the pairs (13) and (24) in the IR's of dimension 6 and 10
(resp. 10 and 6) and to one the IR's of dimension 20 with both (13) and (24) in
the IR of dimension 10. Several pictures of the ground state, all useful for
some purpose, can be given.
The first one is the fermionic representation of Ref.\cite{li} and corresponds
to the linear combination of all possible configurations with all 4 sites 
different, the relative coefficients being the sign of the permutation. 
One can also write this wavefunction as an antisymmetric combination of spin
$SU(2)$ singlets along the horizontal bonds times orbital $SU(2)$ singlets
along the vertical bonds minus the bond exchanged state 
(see fig. \ref{singlet}).
\begin{figure}
\centerline{\psfig{file=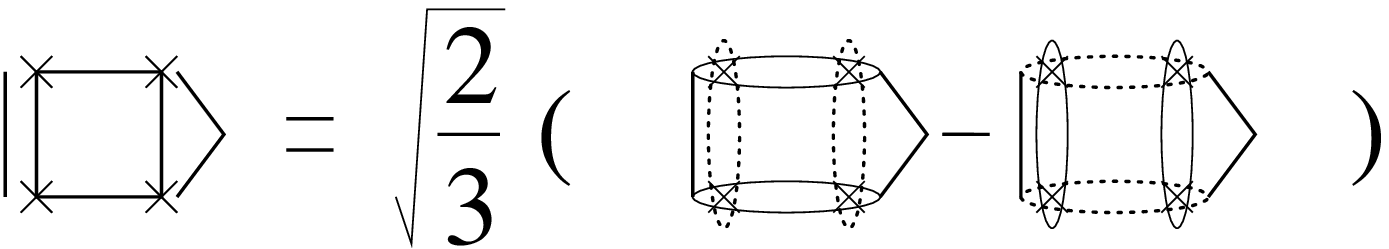,height=1.5cm,angle=0}}
\label{singlet}
\caption{The $SU(4)$ singlet on a four site cluster in terms of spin (solid)
and orbital (dotted) $SU(2)$ singlets}
\label{singlet}
\end{figure}
Finally, if one considers the 4-site cluster as two coupled pairs, the
ground state can be written in terms of pair ground state
only. Since the ground state minimizes the energy of each bond,
this means that the energy of the bonds that couple the pairs is
completely recovered by lifting the degeneracy of the ground state manifold of
two independent pairs. This is another way to understand why the plaquettes play
such a special role.
\begin{figure}[h]
\centerline{\psfig{file=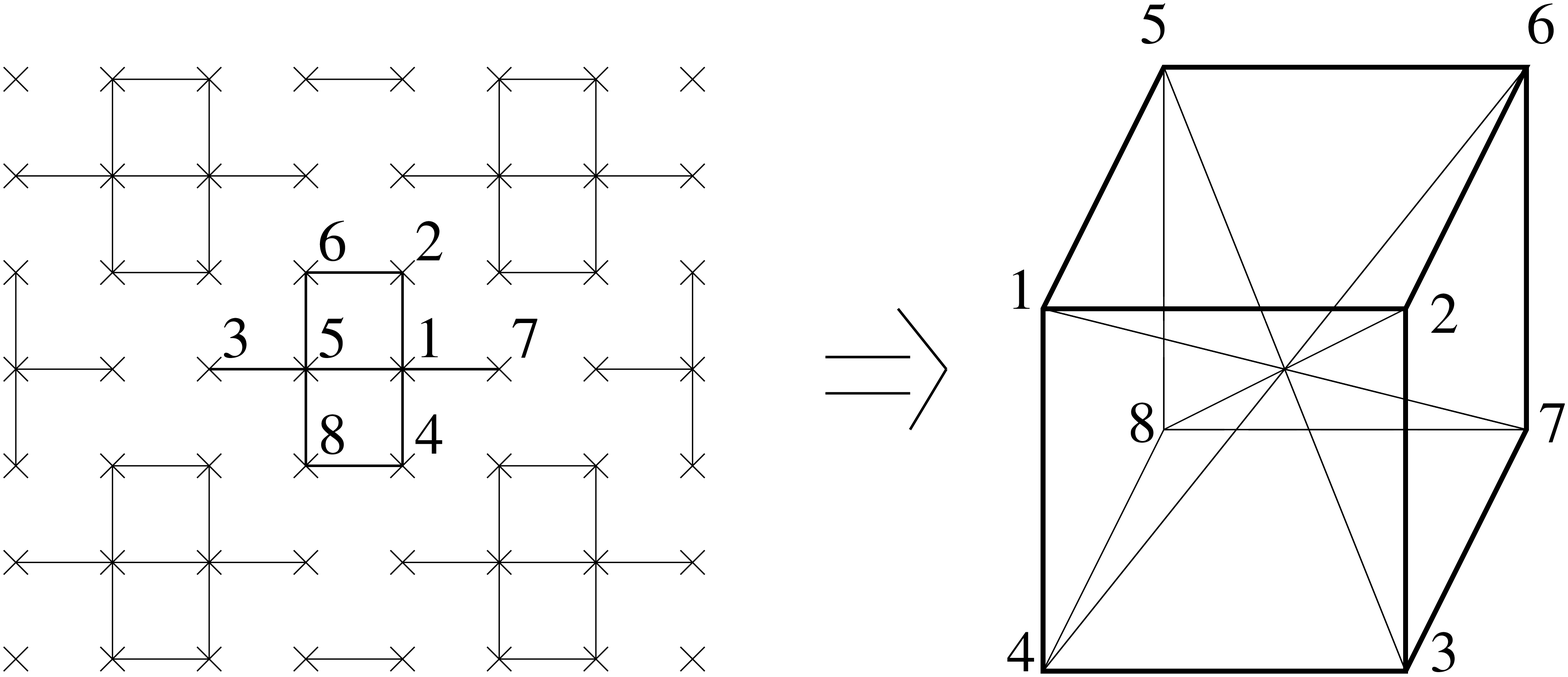,height=4.cm,angle=0}}
\caption{Embedding of the 8-site cluster in the square lattice,
and equivalent connectivity to a cube with diagonals. }
\label{cluster8}
\end{figure}

{\it iii) Eight sites:} The eight-site cluster with periodic boundary conditions
has the topology of a cube with diagonals (see fig. \ref{cluster8}). This allows one
to write the Hamiltonian as
\begin{equation}
H = 4J (C_{1...8} - C_{1368} - C_{2457}+1)
\end{equation}
The ground state is a four-fold degenerate $SU(4)$ singlet of energy -8$J$ 
with the sets (1368) and (2457) in the IR of dimension 20 
which is realized twice for 4 sites\cite{li} and has a Casimir equal to 3/2. 
The first excited singlet 
has energy -4$J$ and is highly degenerate. It is above the first multiplet 
($-6J$), so that there are exactly 4 singlets below the first multiplet.

\begin{figure}
\centerline{\psfig{file=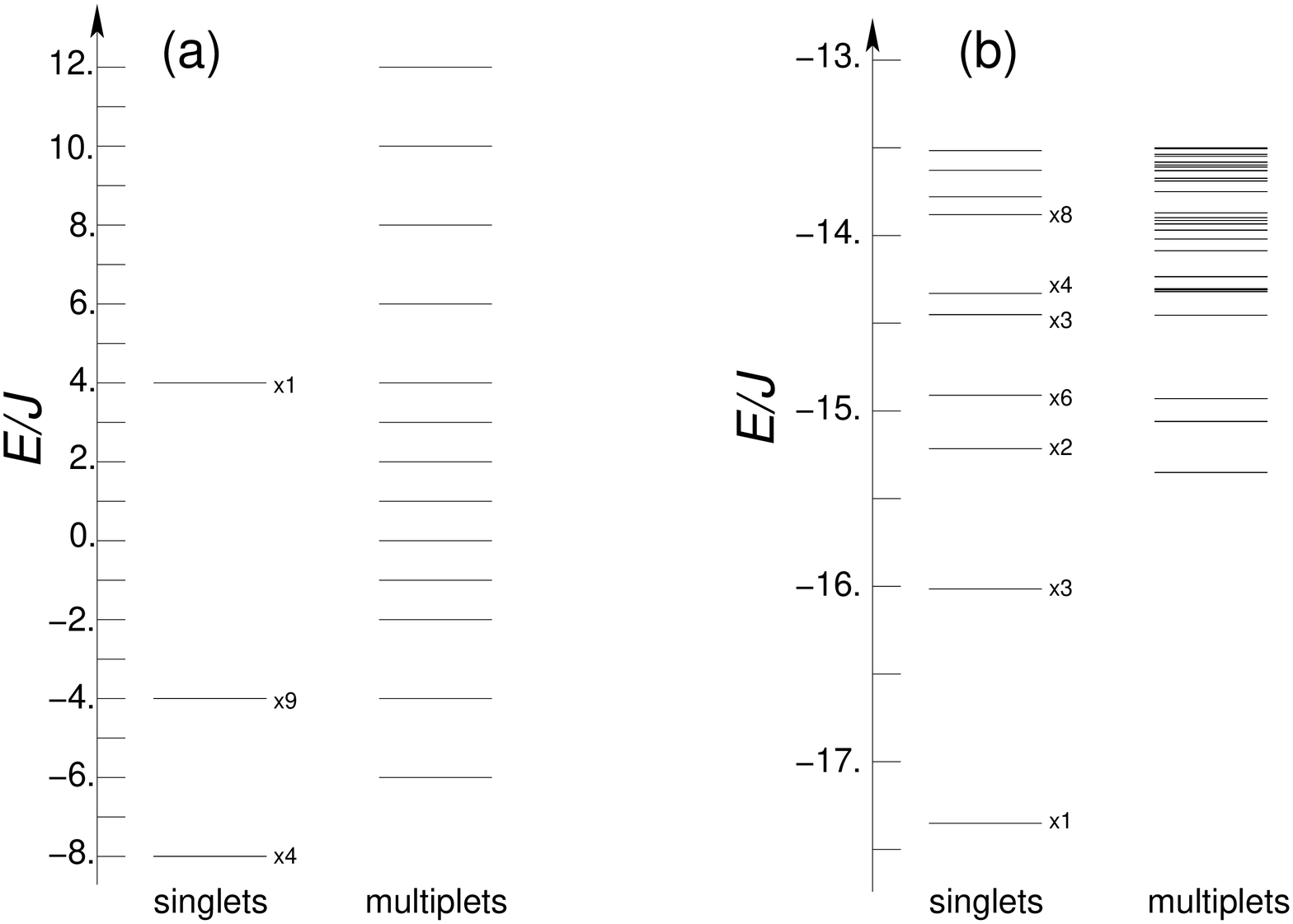,height=6.5cm,angle=0}}
\caption{
Distribution of the low-lying states in the singlet vs. multiplet
representations of $$SU(4)$$ on the 8-site cluster (a) and on the 16-site cluster (b). 
The degeneracy is indicated for the low-lying singlets.}
\label{spectrum}
\end{figure}

\begin{figure}
\centerline{\psfig{file=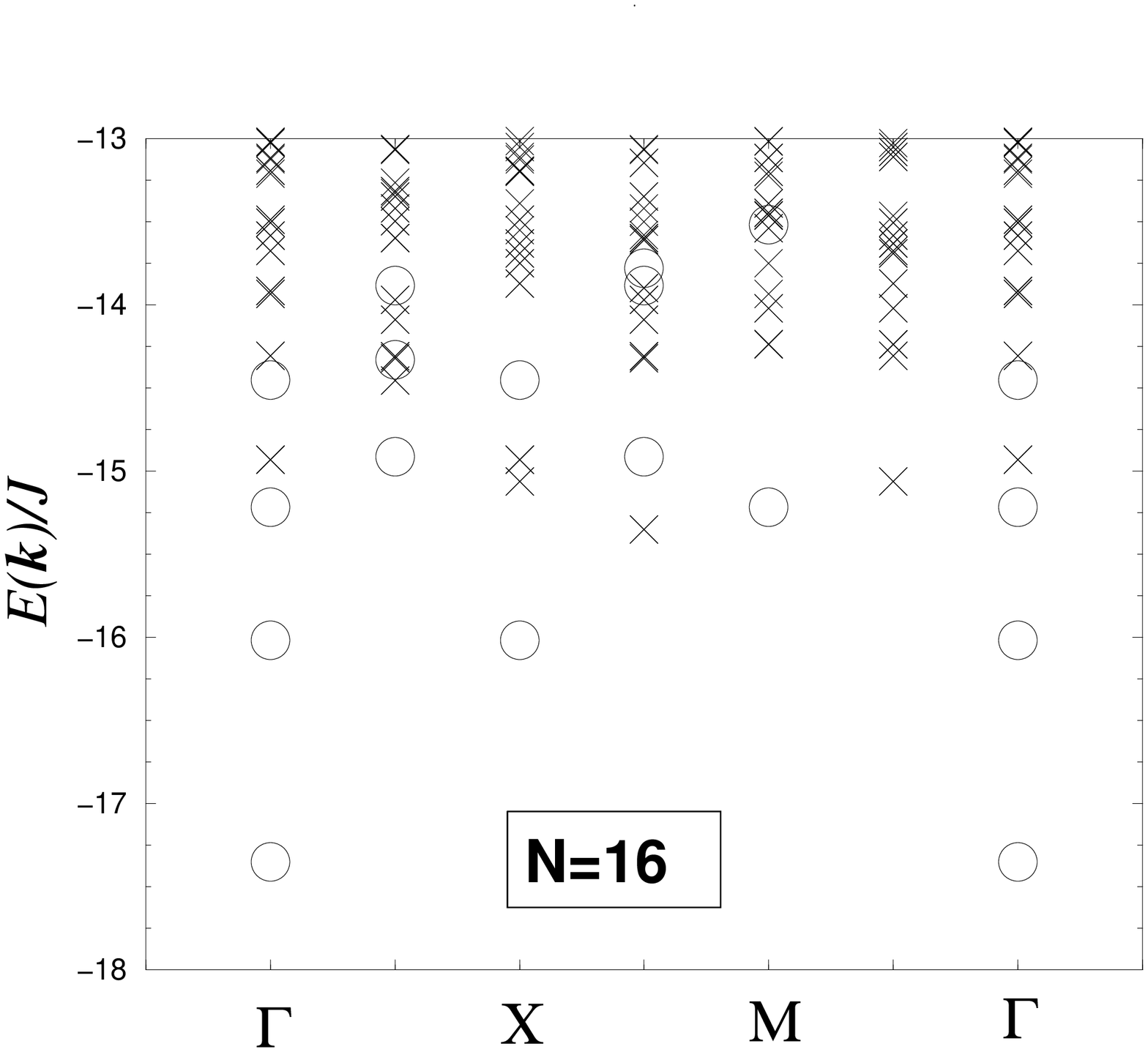,height=7.5cm,angle=-90}}
\caption{Dispersion of low-lying states of the $N=16$ $$SU(4)$$ 
spin-orbital model on a square lattice, through the 6 non-equivalent 
points of the first Brillouin zone. Circles stand for $SU(4)$ singlet states,
and crosses for $SU(4)$ multiplets.}
\label{disp16}
\end{figure}

{\it iv) Sixteen sites:} For that cluster, the only way to get the spectrum is
to perform exact diagonalizations. With 4 degrees of freedom per-site, the
numerical task is roughly equivalent to 32 sites for spin 1/2, and using the
current facilities, this is the largest cluster we could do. To reduce the 
size of the Hilbert space, we used the 3 $SU(4)$ quantum numbers as well as
spatial symmetries. The results are
given in figs. \ref{spectrum}.b. and \ref{disp16}. The ground state is a non-degenerate singlet of energy
-17.351 $J$. The first excitations are singlets. There are three
singlet excitations below the  first multiplet. The dispersion of the singlets
shows a clear
minimum at the X-point, i.e $\vec k =$ (0,$\pi$) and ($\pi$,0).

Let us now discuss these results. The first striking feature is that some basic
quantities have a very small size dependence between 8 and 16 sites\cite{note1}. For
instance, the ground-state energy per site is -$J$ for 8 sites and -1.084425 $J$ for
16 sites. More interestingly, the singlet-multiplet gap is equal to 2$J$ for 8
sites and 1.999809 $J$ for 16 sites (see fig. \ref{spectrum}). This is a clear evidence that
there is a very short
correlation length in the system. Besides, although we have results for only
two different-size systems, the fact that the singlet-multiplet gap is
almost a constant within $10^{-4}$ strongly suggests that it will remain finite
in the thermodynamic limit. This suggests that the ground state does not have 
long-range order, and we must be dealing with some kind of spin-orbital liquid.

To characterize this spin-orbital liquid, we have analyzed in more details 
the singlet sector. Let us start with the 8-site cluster. The ground state
energy is twice the energy of a 4-site plaquette. This suggets that the
plaquette picture of Li {\it et al.}\cite{li} should be a very good starting point. In
an 8-site cluster, there are  18 plaquette coverings, which generate a
Hilbert space of dimension 14. This is identical to the dimension of
the singlet subspace as we checked by diagonalizing the
total Casimir operator. So the ground-state must be a linear combination of the
plaquette coverings. Note that a single plaquette covering is not a ground
state of the Hamiltonian: Using the  fermionic representation of the plaquette
ground state, one can easily check that the mean value of the Hamiltonian in a
plaquette covering is only -6$J$, quite far from the ground-state energy
-8$J$.  So the energy gained through the resonance between plaquette
coverings is crucial to get -$J$ per site in the ground state.
In other words, the exact ground state of the eight-site cluster
is a resonant plaquette state, a generalization of the resonant-valence bond
studied in the context of high temperature superconductivity.
For the 16-site cluster, this picture turns out to remain very accurate. 
In that case, there are 30 plaquette coverings, which is much smaller than 
the dimension of the total singlet subspace (24024). 
To check whether this subspace provides a
good variationnal basis for the ground state, we have numerically calculated 
the projection of the ground state on an orthonormalized basis of this resonant
plaquette subspace, and we find that the weight
of the ground-state in that subspace is 93.5\% of the total weight. In view of
the small relative size of this resonant plaquette subspace, this number is
quite impressive.

\begin{figure}
\centerline{ \psfig{file=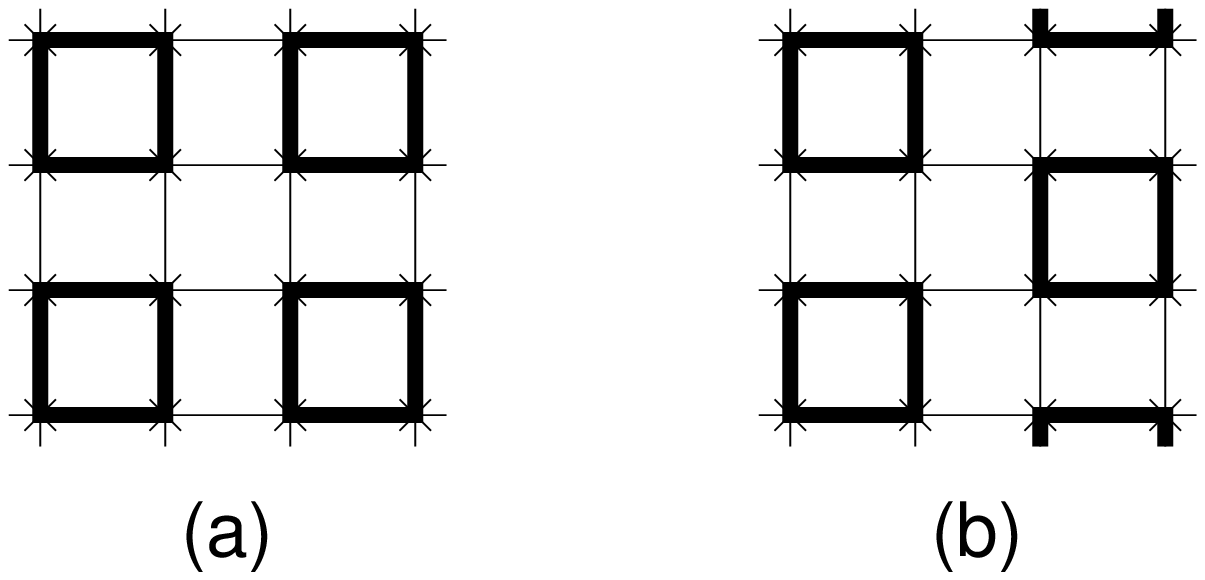,height=4cm,angle=0}}                      
\caption{Two different kinds of plaquette coverings on the square lattice:
(a) A symmetric square lattice of plaquettes. (b) A column (shifted) 
covering.}
\label{coversqr}
\end{figure}    

Another argument in favour of this resonant plaquette description is provided
by the dispersion of the 16-site cluster. First of all, we note that 30\% of
the weight is actually carried by the 4 symmetric plaquette coverings. They
correspond to the plaquette covering of fig. \ref{coversqr}.a., and the three
coverings deduced by translation.  Now the natural singlet excitations from such a
state consist of shifting either rows  or columns of plaquettes (see fig.
\ref{coversqr}.b.). Note however that it is  impossible to do both
simultaneously. So we expect to find quasi-1D low-lying singlet excitations along 
(0,$\pi$) and ($\pi$,0). This is precisely the case for the 16-site cluster
shown in our numerical calculations, with a clear local minimum at the X-point for
singlet excitations.

\begin{figure}
\centerline{ \psfig{file=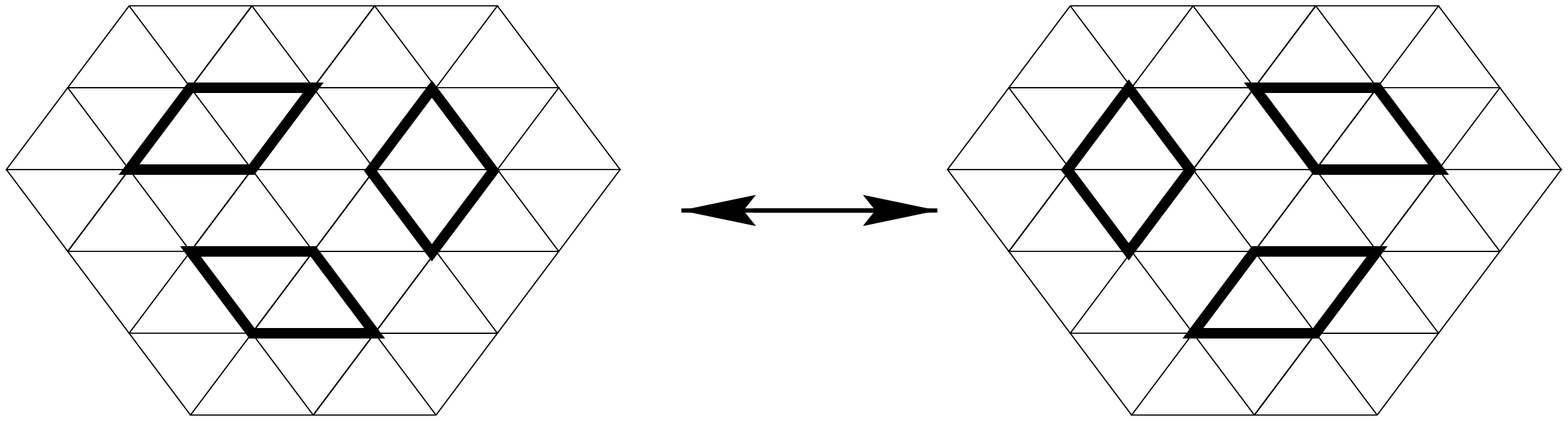,height=2.2cm,angle=0}}
\caption{ Two plaquette coverings of the triangular lattice obtained
by local permutation of three plaquettes.}
\label{permutri}
\end{figure}

Since we have good evidence that a singlet-multiplet gap will remain in the
thermodynamic limit, it is natural to ask whether singlet excitations will
remain within this gap in the thermodynamic limit, as for instance in the case
of the $SU(2)$ antiferromagnet on the Kagom\'e lattice\cite{lecheminant}. 
While we cannot decide
whether some singlets will remain within the gap, it is quite unlikely that 
they will form a continuum, like for the Kagom\'e lattice: The number of 
low-lying singlets does not increase from 8 to 16 sites - it is equal to 4 in
both cases - and they are shifted to higher energy on going form 8 to 16 sites.
This result might be explained by a simple counting argument. Because of the
impossibility to shift simultaneously rows and columns, the number of
plaquette coverings of the square lattice does not increase exponentially with
the size of the system, but with the square root of the size. So we do not
expect to be able to build a continuum with a number of states that remains
significant in the thermodynamic limit. This should be contrasted to the Kagom\'e
case, where a simple counting argument could reproduce the exponential increase
of the number of low-lying singlets\cite{mila}. In that respect, 
the $SU(4)$ model on the
triangular lattice might be different. In that case, it is possible to generate
new plaquette coverings by local modifications (see fig. \ref{permutri}), and 
the number of
plaquette coverings increases exponentially with the number of sites. So, if plaquette
coverings provide the lowest singlet states for the triangular lattice,
a continuum of
low-lying $SU(4)$ singlets might be present in the singlet-multiplet gap. This
would be consistent with the specific heat data of LiNiO$_2$, where no gap 
was observed\cite{kitaoka}.

Finally, let us comment on the physical properties of Mott insulators
that can be described by a 2D spin-orbital model in the vicinity of the $SU(4)$
symmetry. The correlation length appears to be so short, and the
singlet-multiplet gap so large, that small perturbations will not 
close the gap.  So in the absence of coupling to other degrees of freedom the
system is expected to remain disordered with a gap to magnetic excitations.
Besides, the  cooperative Jahn-Teller mechanism that lifts the orbital
degeneracy will also have to overcome this singlet-multiplet gap since the
$SU(4)$ singlet ground-state of a plaquette explicitly requires two orbital
degrees of freedom (see fig. \ref{singlet}).  As a consequence, the orbitals will not
order unless the electron-phonon coupling is strong enough. This picture is
consistent with the properties reported so far for LiNiO$_2$\cite{kitaoka}. As we noticed
above, the structure of the singlet sector might be different for the
triangular lattice, however. The analysis of this model is under progress.

We would like to aknowledge 
useful discussions with Karlo Penc, Beat Frischmuth, Anatoli Stepanov and Nicolas Destainville.
The numerical simulations were performed on the Cray superconputers of the IDRIS
(Orsay, France). This work was supported in part by DOE Grant No.
DE/FG03-98ER45687.

\end{multicols}


\begin{thebibliography}{0}
\bibitem{note1} We have not included the 4-site cluster in this comparison
because it has only one bond per site, whereas the 8 and 16 site clusters have
two.
\end{thebibliography}
\end{document}